\documentclass{scrartcl}
\usepackage[pdftex]{graphicx}
\usepackage{tikz}
\usepackage{amsmath}
\usepackage{amsfonts}
\usepackage{amssymb}

\newtheorem{thm}{Theorem}[section]

\usetikzlibrary{calc}

\usepackage{url}

\begin{document}
%
\title{A Game Theoretical Analysis of Localization Security
  in Wireless Sensor Networks with Adversaries}
 
\def\IEEEauthorblockN#1{#1\\}
\def\IEEEauthorblockA#1{#1}
\def\IEEEpeerreviewmaketitle{}

\author{\IEEEauthorblockN{Nicola Gatti}
\IEEEauthorblockA{Politecnico di Milano, DEI\\
P.za L. da Vinci 32, I--20133 Milan, Italy \\
Email: \url{ngatti@elet.polimi.it}}
\and
\IEEEauthorblockN{Mattia Monga}
\IEEEauthorblockA{Universit\`a degli Studi di Milano, DICo\\
Via Comelico 39, I--20135 Milan, Italy\\
Email: \url{mattia.monga@unimi.it}}
\and
\IEEEauthorblockN{Sabrina Sicari}
\IEEEauthorblockA{Universit\`a dell'Insubria, DICOM\\
Via Mazzini 5, I--21100 Varese, Italy\\
Email: \url{sabrina.sicari@uninsubria.it}}}


\maketitle

\begin{abstract}
Wireless Sensor Networks (WSN) support data collection and distributed
data processing by means of very small sensing devices that are easy
to tamper and cloning: therefore classical security solutions based on access control and strong authentication are
difficult to deploy. In this paper we look at the problem of
assessing security of node localization. In particular, we analyze the scenario in which Verifiable
Multilateration (VM) is used to localize nodes and a malicious node (i.e., the adversary) try to masquerade as non-malicious. We resort to non-cooperative game theory and we model this scenario as a two-player game. We analyze the optimal players' strategy  and we show that the VM is indeed a proper mechanism to reduce fake positions.
\end{abstract}


%
\IEEEpeerreviewmaketitle

\section{Introduction}
\label{sec:introduction}

Wireless Sensor Networks (WSN)~\cite{Akyildiz02,baronti:07}
technologies support data collection and distributed data processing
by means of very small sensing devices. Nowadays, sensors are used in
many contexts such as surveillance systems, systems supporting traffic
monitoring and control in urban/suburban areas, military and/or
anti-terrorism operations, telemedicine, assistance to disabled and
elderly people, environmental monitoring, localization of services and
users, and industrial process control.  This activities rely greatly on
data about the positions of sensor nodes.  Nodes are often deployed
randomly or move, and one of the challenges is computing localization
at time of operations. Several localization approaches have been
proposed (for example, \cite{Estrin00, capkun02, chen02, Doherty01,
  He, Nicu, Rama, Savvides}), but most of the current approaches omit
to consider that WSNs could be deployed in an adversarial setting,
where hostile nodes under the control of an attacker coexist with
faithful ones.  In fact, wireless communications are easy to tamper
and nodes are prone to physical attacks and cloning: thus classical
solutions, based on access control and strong authentication, are
difficult to deploy.

An approach to localize nodes even when some of them are compromised
was proposed in \cite{capkun} and it is known as \textit{Verifiable
  Multilateration} (VM). However, in some situations also using
Verifiable Multilateration the security localization behavior of a
node is undefined, in other words there is not enough information for
considering it a secure or malicious node. This weakness could be
exploited by a malicious node to masquerade as an undefined one,
pretending to be in a position that is still compatible with all
verifiers' information. To the best of our knowledge, the analysis of
this scenario has not been explored so far in the literature: we
explicitly consider how a malicious node, on the one side, could act
and, on the other side, how the system could face it. This constitutes
the original contribution of our work.

In this paper, we resort to non-cooperative game theory to study our scenario. More precisely, we model it as a two-player strategic-form game, where the first player is a \emph{verifier} that uses VM and the second player is a \emph{malicious node}. The verifier acts  to securely localize the malicious node, while the malicious node acts  to masquerade as undefined.  As is customary in game theory, the players are considered rational (i.e., maximizers). This amounts to say that the malicious node is modeled as the strongest adversary. We study the game, showing some results concerning the robustness of VM.  The paper is organized as follows: Section~\ref{sec:VM} provides a short overview about Verifiable Multilateration; Section~\ref{sec:secure-game} shortly describes secure localization game, providing some basic concepts; Section~\ref{sec:strategic} introduces strategic game analysis. Section~\ref{sec:conclusion} draws some conclusions and provides hints for future works.


\section{Verifiable Multilateration}\label{sec:VM}
Multilateration is a technique used in WSNs to
estimate the coordinates of the unknown nodes, given the positions of
some given landmark  nodes, called \emph{anchor} nodes, whose positions
are known. The position of the unknown node $U$ is computed by
geometric inference based on the distances between the anchor
nodes and the node itself. However, the distance is not measured
directly; instead, it is derived by knowing the speed of the
signal in the medium used in the transmission, and by measuring
the time needed to get an answer to a beacon message sent to $U$.

\begin{figure}
  \centering
   \begin{tikzpicture}
            \coordinate [label=below:$v_1$] (V1) at (0,0);
            \coordinate [label=below:$v_2$] (V2) at (50pt,0);
            \path let
             \p1 = (V2),
             \n{h} = {sqrt(veclen(\p1)^2-25pt^2)}
            in coordinate [label=above:$v_3$] (V3) at (25pt,\n{h});
            \coordinate [label=below:$u$] (U) at (25pt,25pt);
            \draw[dashed] (V1) -- (V2) -- (V3) -- cycle;
            \clip (U) circle(30pt);
            \foreach \i in {1, 2, 3} {
              \draw [red] (V\i) -- node[black] {\tiny $db_\i$} (U);
              \draw let
              \p\i = ($ (U) - (V\i) $),
              \n\i = {veclen(\p\i)}
              in (V\i)
              circle(\n\i);
            }
          \end{tikzpicture}

  \caption{Verifiable multilateration}
  \label{fig:vm}
\end{figure}
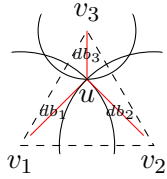

Unfortunately, if this computation is carried on without any
precaution, $U$ might fool the anchors by delaying the beacon
message. However, since a malicious node can delay the answer
beacon, but not speed it up, under some conditions it is possible
to spot malicious behaviors. VM uses
three or more anchor nodes to detect misbehaving nodes. In VM the
anchor nodes work as \emph{verifiers} of the localization data and
they send to the sink node $B$ the information needed to evaluate
the consistency of the coordinates computed for $U$. The basic
idea of VM is shown in Figure~\ref{fig:vm}: each verifier $V_i$
computes its \emph{distance bound}~\cite{Brands} to $U$; any point
$P \neq U$ inside the triangle formed by $V_1,V_2,V_3$ has
necessarily at least one of the distance to the $V_i$ enlarged.
This enlargement, however, cannot be masked by $U$ by sending a
faster message to the corresponding verifier.

Under the hypothesis that verifiers are trusted and they can
securely communicate with $B$, the following verification process
can be used to check the localization data:

\begin{enumerate}
\item Each verifier $V_i$ sends a beacon message to $U$ and
records the
  time $\tau_i$ needed to get an answer;
\item Each verifier $V_i$ (whose coordinates $\langle x_i, y_i
  \rangle$ are known) sends to $B$ a message with its $\tau_i$;
\item From $\tau_i$, $B$ derives the corresponding distance bound
  $db_i$ (that can be easily computed if the speed of the signal is
  known) and it estimates $U$'s coordinates by minimizing the sum of
  squared errors
  \[ \epsilon = \sum_i (db_i - \sqrt{(x - x_i)^2 + (y -
    y_i)^2})^2 \] where $\langle x,y \rangle$ are the (unknown)
  coordinates to be estimated\footnote{In an ideal situation where
    there are no measurement errors and/or malicious delays this is
    equivalent to finding the (unique) intersection of the circles
    defined by the distance bounds and centered in the $V_i$ (see
    Figure~\ref{fig:vm}) and $\epsilon = 0$.};
\item $B$ can now check if $\langle x, y \rangle$ are feasible in
the given
  setting by two incremental tests:
(a)~\emph{$\delta$-test:} For all verifiers $V_i$, compute the
distance between the estimated $U$ and $V_i$: if it differs from
the measured distance bound by more than the expected distance
measurement error, the estimation is affected by malicious
tampering; (b)~\emph{Point in the triangle test:} Distance
bounds are reliable
  only if the estimated $U$ is
  within at least one verification triangle formed by a triplet of
  verifiers, otherwise the estimation is considered unverified.

\end{enumerate}

If both the \emph{$\delta$} and the \emph{point-in-the-triangle}
tests are positive, the distance bounds are consistent with the
estimated node position, which moreover falls in at least one
verification triangle. This means that none of the distance bounds
were enlarged. Thus, the sink can consider the estimated position
of the node as $\textsc{Robust}$; else, the information at hands is
not sufficient to support the reliability of the data. An
estimation that does not pass the $\delta$ test is considered
$\textsc{Malicious}$. In all the other cases, the sink marks the
estimation as $\textsc{Unknown}$. In an ideal situation where there
are no measurement errors, there are neither malevolent nodes
marked as $\textsc{Robust}$, nor benevolent ones marked as
$\textsc{Malicious}$. Even in this ideal setting, however, there are
$\textsc{Unknown}$ nodes, that could be malevolent or not. In other
words there are no sufficient information for evaluating the
trustworthiness of node position. In fact, $U$ could pretend, by an
opportune manipulation of delays, to be in a position $P$ that is
credible enough to be taken into account. No such points exist
inside the triangles formed by the verifiers (this is exactly the
idea behind verifiable multilateration), but outside them some
regions are still compatible with all the information verifiers
have.

Consider $N$ verifiers that are able to send signals in a
range $R$. Let $x_0$ and $y_0$ the \emph{real}
coordinates of $U$. They are unknown to the verifiers, but
nevertheless they put a constraint on plausible fake positions,
since the forged distance bound to $V_i$ must be greater than the
length of $\overline{UV_i}$.

Thus, any point $P=\langle x,y\rangle$ that is a plausible
falsification of $U$ has to agree to the following constraints, for each
$1 \leq i \leq N$:

\begin{equation}
  \left\{
    \begin{array}{l}
      {\left(y - y_{i}\right)}^{2} + {\left(x - x_{i}\right)}^{2} <
      R^{2} \\
{\left(y - y_{i}\right)}^{2} + {\left(x - x_{i}\right)}^{2} > {\left(y_{0} - y_{i}\right)}^{2} + {\left(x_{0} - x_{i}\right)}^{2}\\

    \end{array} \right.
  \label{eq:constraints}
 \end{equation}

 The constraints in (\ref{eq:constraints}) can be understood better by
 looking at Figure~\ref{fig:plausible}, where three verifiers are
 depicted: the green area around each verifier denotes its power
 range, and the red area is the bound on the distance that $U$ can put
 forward credibly. Thus, any plausible $P$ must lay outside every red
 region and inside every green one.

\newcommand{\Rval}{150}
\newcommand{\xival}{0}
\newcommand{\yival}{0}
\newcommand{\xiival}{100}
\newcommand{\yiival}{0}
\newcommand{\xiiival}{50}
\newcommand{\yiiival}{100}
\newcommand{\xzval}{20}
\newcommand{\yzval}{20}

\begin{figure}[h]

  \centering

  \begin{tikzpicture}[scale=.28, text opacity=1]
  \coordinate [label=below:$V_1$] (V1) at (\xival pt,\yival pt);
  \coordinate [label=below:$V_2$] (V2) at (\xiival pt,\yiival pt);
  \coordinate [label=above:$V_3$] (V3) at (\xiiival pt,\yiiival pt);
  \coordinate [label=above:$U$] (U) at (\xzval pt,\yzval pt);
  \filldraw (U) circle (2pt);
  \coordinate [label=above:$P$] (P) at (-40 pt,20 pt);
  \filldraw (P) circle (1pt);
  \draw [dotted] (V1) -- (V2) -- (V3) -- cycle;
  \foreach \i in {1, 2, 3}
  { \filldraw[dashed, fill=red, fill opacity=.4] let
              \p\i = ($ (U) - (V\i) $),
              \n{d\i} = {veclen(\p\i)}
            in
              coordinate (p\i) at (\p\i) (V\i)  circle (\n{d\i});
    \filldraw[dashed, fill=green, fill opacity=.1] (V\i)  circle
    (\Rval pt);
  }

\end{tikzpicture}
  \caption{Plausible falsification region: $P$ is a plausible fake
    position for $U$ since lays outside every red region and inside
    every green one (and it is outside the triangle of verifiers).}
\label{fig:plausible}
\end{figure}
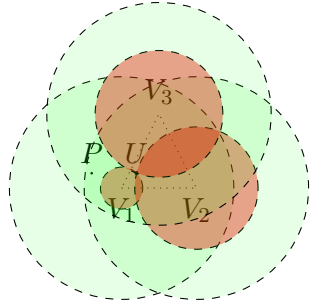



\section{Secure localization game}
\label{sec:secure-game}

Our aim is the study of the behavior of a possible malicious node that
acts to masquerade as an unknown node and, at the same time, how the
malicious node can be faced at best by the verifiers. This is a
typical non-cooperative setting that can be analyzed by leveraging on
game theoretical models. A \emph{game} is described by a couple: \emph{mechanism} and \emph{strategies}. The mechanism defines the rules of the game in terms of number of players and actions available to the players. The strategies describe the behaviors of the players during the game in terms of played actions. Strategies can be pure, when a player acts one action with a probability of one, or they can be mixed, when a player randomizes over a set of actions. The players' strategies define an outcome (if the strategies are pure) or a randomization over the outcomes (if mixed). Players have preferences over the outcomes expressed by utility functions and each player is \emph{rational}, acting to maximize its own utility. Solving a game means to find a profile of strategies (i.e., a set specifying one strategy for each player) such that the players'
strategies are somehow in equilibrium. The most known equilibrium
concept is \emph{Nash} where each player cannot improve its own
utility by deviating unilaterally (a detailed treatment of Nash
equilibrium can be found in \cite{osborne94}): a fundamental result in
the study of equilibria is that every game admits at least one Nash equilibrium in mixed strategies, while pure strategy equilibrium might not exist.

We now formally state our secure localization game, by focusing on a
setting with $N=3$ verifiers. It is a tuple $\langle Q, A, u
\rangle$. Set $Q$ contains the  players and is defined as
$Q=\{\mathbf{v}, \mathbf{m}\}$ ($\mathbf{v}$ denotes the verifiers and
$\mathbf{m}$ denotes the malicious node). Set $A$ contains the players
actions. More precisely, given a surface $S\subseteq \mathbb{R}^2$,
the actions available to $\mathbf{v}$ are all the possible tuples of
positions $\langle V_1, V_2, V_3 \rangle$ of the three verifiers with
$V_1, V_2, V_3 \in S$, while the actions available to $\mathbf{m}$ are
all the possible couples  of positions $\langle U,P\rangle$ with $U,P
\in S$ (where $U$ and $P$ are defined in the previous section). We
denote by $\sigma_{\mathbf{v}}$ the strategy (possibly mixed) of
$\mathbf{v}$ and by $\sigma_{\mathbf{m}}$ the strategy (possibly
mixed) of $\mathbf{m}$. Given a strategy profile
$\sigma=(\sigma_{\mathbf{v}}, \sigma_{\mathbf{m}})$ in pure strategy,
it is possible to check whether or not constraints
(\ref{eq:constraints}) are satisfied. The outcomes of the game can be
$\{\textsc{malicious}, \textsc{robust}, \textsc{unknown}\}$. Set $u$
contains the players' utility functions, denoted
$u_{\mathbf{v}}(\cdot)$ and $u_{\mathbf{m}}(\cdot)$ respectively, that
define their preferences over the outcomes. We define
$u_{i}(\textsc{malicious})=u_{i}(\textsc{robust})=0$ for $i\in
\{\mathbf{v}, \mathbf{m}\}$, while $u_{i}(\textsc{unknown})$ can be
defined differently according to different criteria. A simple
criterion could be to assign $u_{\mathbf{v}}(\textsc{unknown})=-1$ and
$u_{\mathbf{m}}(\textsc{unknown})=1$. However, our intuition is that the $\textsc{unknown}$ outcomes are not the same for the players, because $\mathbf{m}$ could prefer those in which the distance between $U$ and $P$ is maximum. In particular we propose three main criteria to characterize $\textsc{unknown}$ outcomes:
\begin{enumerate}
\item \emph{maximum deception}, $u_{\mathbf{m}}$ is defined as the distance between $U$ and $P$, while $u_{\mathbf{v}}$ is defined as the opposite; 
\item \emph{deception area}, $u_{\mathbf{m}}$ is defined as the size of the region $S'\subseteq S$ such that $P\in S'$ is marked as \textsc{unknown}, while $u_{\mathbf{v}}$ is defined as the opposite; 
\item \emph{deception shape}, $u_{\mathbf{m}}$ is defined as the number of disconnected regions $S'\subseteq S$ such that $P\in S'$ is marked as \textsc{unknown}, while $u_{\mathbf{v}}$ is defined as the opposite. 
\end{enumerate}
Players could even use different criteria, e.g., $\mathbf{v}$ and
$\mathbf{m}$ could adopt the maximum deception criterion and the
deception shape respectively. However, when players adopt the same
criterion, the game is \emph{zero-sum}, the sum of the players'
utilities being zero. This class of games is easy and has the property
that the maxmin, minmax, and Nash strategies are the same. In this
case calculations are simplified by the property that $u_{\mathbf{v}}
=-u_{\mathbf{m}}$; in the following we shall adopt this assumption. 


\section{Game Analysis}
\label{sec:strategic}

For the sake of simplicity, we focus on the case in which  both
players adopt the maximum deception criterion. In principle, however, our
analysis can be extended to other criteria: in particular,
Theorem~\ref{thm:primo} is valid for all the proposed criteria.

\subsection{Analysis with Pure Strategies}
\label{subsec:pure}

In this section, we show that there can be no equilibrium in pure
strategies. We discuss also what is the value of the maximum deception
when the verifiers adopts a pure strategy. We consider only the case
in which $\forall i,j\,\overline{V_iV_j}\leq R$ since otherwise the
region in which VM would be applicable is small and no
\textsc{unknown} positions would be possible, thus paradoxically the
verifiers would have an incentive to reduce it further to only one point,
making the localization procedure worthless.

At first, we can show that for each action of the verifiers, there
exists an action of the malicious node such that this is marked as
\textsc{unknown}. 


\begin{thm}
For each tuple $\langle V_1, V_2, V_3\rangle$ such that $\overline{V_iV_j}\leq R$ for all $i,j$, there exists at least a couple $\langle U,P\rangle$ such that $u_{\mathbf{m}}>0$. \label{thm:primo}
\end{thm}

\emph{Proof.} Given $V_1, V_2, V_3$ such that $\overline{V_iV_j}\leq
R$ for all $i,j$, choose a $V_i$ and call $X$ the point on the line
$\overline{V_kV_j}\;(k,j\neq i)$ closest to $V_i$. Assign $U=X$. Consider the line connecting $V_i$ to $X$, assign $P$ to be any point $X'$ on this line such that $\overline{V_iX}\leq \overline{V_iX'}\leq R$. Then, by construction $u_{\mathbf{m}}>0$.\hfill$\Box$

We discuss what is the configuration of the three verifiers, such that the maximal deception is minimized.

\begin{thm}
Any tuple $\langle V_1, V_2, V_3\rangle$ such that $\overline{V_iV_j}= R$ for all $i,j$ minimizes the maximum deception.\label{thm:triangoloequilatero}
\end{thm}

\emph{Proof.} Since we need to minimize the maximum distance between
two points, by symmetry, the triangle whose vertexes are $V_1, V_2,
V_3$ must have all the edges with the same length. We show that $\overline{V_iV_j}=R$. It can easily seen, by geometric construction, that $U$ must be necessarily inside the triangle. As shown in Section~2, $P$ must be necessarily outside the triangle and, by definition, the optimal $P$ will be on the boundary constituted by some circle with center in a $V_i$ and range equal to $R$ (otherwise $P$ could be moved farther and $P$ would not be optimal). As $\overline{V_iV_j}$ decreases, the size of the triangle reduces, while the boundary keeps to be the same, and therefore $\overline{UP}$ does not decrease.\hfill$\Box$ 

We are now in the position to find the maxmin value (in pure strategies) of the verifiers, i.e., the action that maximizes the verifiers' utility given that the malicious node will minimize it. The problem of finding the maxmin strategy can be formulated as the following non-linear optimization problem:

\begin{eqnarray*}
\max_{\textnormal{constraints (1)}}\overline{UP} & \begin{split}\textnormal{for some $V_1, V_2, V_3$ with} \\ \overline{V_iV_j}= R \textnormal{ for all } i,j\end{split}
\end{eqnarray*}

We solved this problem by using conjugated subgradients. We report the
solution. Called $W$ the orthocenter of the triangle, $U$ and $P$ can
be easily expressed with polar coordinates with origin in $W$. We
assume that $\theta=0$ corresponds to a line connecting $W$ to a
$V_i$. We have, $U = (\rho = 0.1394 R, \theta = \frac{\pi}{6})$ and
$P=(\rho= 0.4286 R, \theta = \frac{\pi}{6}+0.2952)$,  and, for
symmetry,  $U = (\rho = 0.1394 R, \theta = -\frac{\pi}{6})$ and
$P=(\rho= 0.4286 R, \theta = -\frac{\pi}{6}-0.2952)$. Therefore, there
are six optimal couples $\langle U,P\rangle$s. In
Figure~\ref{fig:bestresponse} depicts the
malicious node's best action, by showing on the right all the
symmetrical positions. The value of $u_{\mathbf{m}}$ (i.e., the maximum deception) is $0.2516 R$. In other words, when the verifiers compose an equilateral triangle, a malicious node can masquerade as unknown and the maximum deception is about $25\%$ of the verifiers' range~$R$.

\newlength{\rpower}
\setlength{\rpower}{.15\columnwidth}

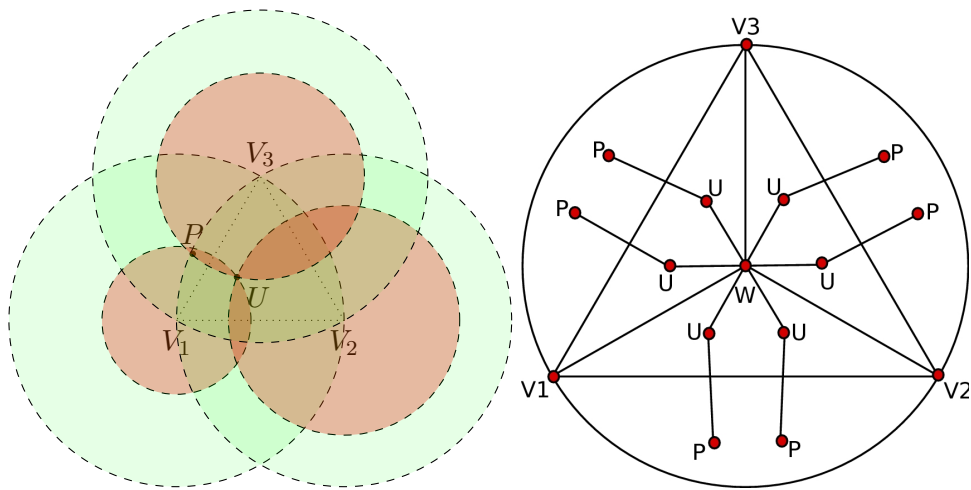
\begin{figure}[h]
  \begin{tikzpicture}[text opacity=1]
  \coordinate [label=below:$V_1$] (V1) at (0 pt, 0 pt);
  \coordinate (H1) at ($(V2)!(V1)!(V3)$);
  \coordinate [label=below:$V_2$] (V2) at ($(V1) + (\rpower,0)$);
  \coordinate (H2) at ($(V1)!(V2)!(V3)$);
  \coordinate [label=above:$V_3$] (V3) at ($(V1)!1!60:(V2)$);
  \coordinate  (W) at (intersection of V1--H1 and V2--H2);

  \coordinate [label=-30:$U$] (U) at ($(W)!.1394\rpower!-30:(V1)$);
  \filldraw (U) circle (1pt);
  \coordinate [label=above:$P$] (P) at ($(W)!.4286\rpower!-46.9137:(V1)$);
  \filldraw (P) circle (1pt);
  \draw [dotted] (V1) -- (V2) -- (V3) -- cycle;
   \foreach \i in {1, 2, 3}
   { \filldraw[dashed, fill=red, fill opacity=.3] let
               \p\i = ($ (U) - (V\i) $),
               \n{d\i} = {veclen(\p\i)}
             in
               coordinate (p\i) at (\p\i) (V\i)  circle (\n{d\i});
     \filldraw[dashed, fill=green, fill opacity=.1] (V\i)  circle
     (\rpower);
  }

\end{tikzpicture}
\includegraphics[width=.4\columnwidth]{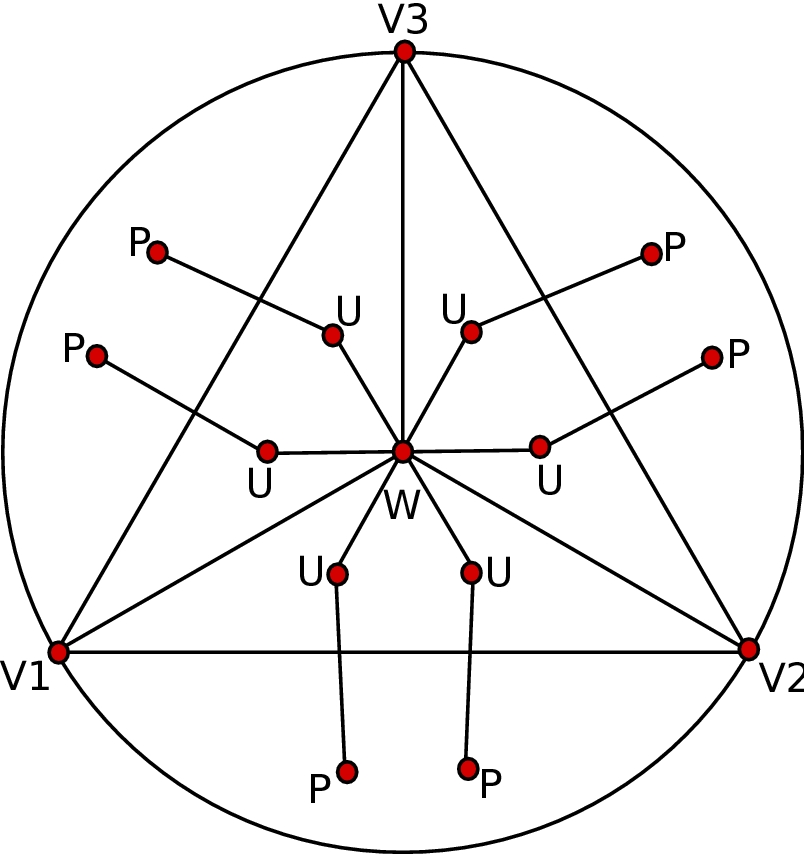}
\caption{Malicious node's best responses.}
\label{fig:bestresponse}
\end{figure}

\begin{figure}[h]
\centering
\end{figure}

We consider the verifiers' strategy and we show that for each action of the malicious node they can find an action such that the malicious node is marked either as \textsc{robust} or as \textsc{malicious}.

\begin{thm}
For each couple $\langle U,P\rangle$, there exists at least a tuple $\langle V_1, V_2, V_3\rangle$ such that $u_{\mathbf{v}}=0$.\label{thm:secondo}
\end{thm}

\emph{Proof}. If $U\equiv W$ (where $W$ is the orthocenter of the equilateral triangle composed by the verifiers), then, by geometric construction, maximum deception is zero (we omit the calculation for reasons of space).\hfill$\Box$

By combining Theorems~\ref{thm:primo} and~\ref{thm:secondo}, we have
that our game cannot admit any Nash equilibrium in pure strategies. Indeed, for each $\sigma_{\mathbf{v}}$ there exists a best response $\sigma_{\mathbf{m}}$ such that $\sigma_{\mathbf{v}}$ is not the best response to $\sigma_{\mathbf{m}}$.

\subsection{Discrete Approximation Hardness}
\label{subsec:hardness}

Finding a mixed strategy equilibrium in a two-player zero-sum finite game is well known to be a polynomial problem in the number of actions available to the players. This is because the problem of finding a minmax strategy can be formulated as a linear mathematical programming problem. However, our problem is not finite, $V_1,V_2,V_3,U,P$ belonging to a continuous space. In this section, we show that finding an approximate solution by discretizing the surface $S$ in a finite number of points is not practically affordable. 

We discretize $S$ by a finite grid with a given step $\Delta$. We call
$S_d\subset S$ the set of points in the grid. The players can choose
their position from set $S_d$. We denote by $A_{\mathbf{v}}$ and
$A_{\mathbf{m}}$ the set of actions of the verifiers and malicious
node respectively. Supposed $S_d$ to be a square and called $l$ the
length of $S$, the number of points in $S_d$ is
$|S_d|=\lceil\frac{l}{\Delta}\rceil^2$. We have that $|A_{\mathbf{v}}|
= \sum_{3 \leq i \leq |V|}{{|S_d|^2}\choose i}\sim O(|S_d|^6)$ and $|A_{\mathbf{m}}|
= |S_d|^2\cdot(|S_d|^2-1)\sim O(|S_d|^4)$. For each possible profile of players' actions we compute $u_{\mathbf{m}}$ as the maximum deception. Notice that the number of all the possible profiles of players' actions is $\sim O(|S_d|^{10})$. We denote by $p_{\mathbf{v}}(i)$ the probability with which $\mathbf{v}$ plays action $i\in A_{\mathbf{v}}$. The linear programming formulation to find the minmax strategy (and equivalently the Nash equilibrium) is:
\begin{eqnarray}
\min u\\
\sum_{i\in A_{\mathbf{v}}} p_{\mathbf{u}}(i) u_{\mathbf{m}}(j,i) \leq u & \forall j\in A_{\mathbf{m}} \label{con1} \\
p_{\mathbf{u}}(i) \geq 0 & \forall i \in A_{\mathbf{v}} \label{con2}\\ 
\sum_{i\in A_{\mathbf{v}}} p_{\mathbf{u}}(i) = 1 & \label{con3}
\end{eqnarray}
Constraints (\ref{con1}) force the expected utility $\mathbf{m}$ receives from taking action $j$ to be not larger than $u$; constraints (\ref{con2}) and (\ref{con3}) grant probabilities $p_{\mathbf{m}}(\cdot)$ to be well defined. The objective function is the minimization of $u$ that by constraints (\ref{con1}) is the maximal expected utility of $\mathbf{m}$.

We solved the above mathematical programming problem with grids with $3,4,5$ points per edge. In all these case studies, the verifiers always mark the malicious node as robust or malicious, and therefore $u_{\mathbf{m}}$ is always equal to zero. We notice that  the utility matrix presents a number of non-null values, anyway, there exists at least a configuration of verifiers such that for no action of the malicious node this is marked as unknown. This is because the grid is too loose. However, with a larger number of points per edge, the problem is not computationally affordable because the number of outcomes is excessively large.

\subsection{Mixed Strategies with a Fixed Orthocenter}
\label{subsec:strategies}

The hardness result discussed in the previous section pushes us to resort to an analytical approach to find the players' equilibrium strategies. Here, we discuss the strategies in a simplified case study. The idea is that this result can provide insight to solve the general case.

At first we show that any equilibrium strategy prescribes that the players randomize over a continuous space of action. Call $supp(\sigma_i)$ the set of actions played with strictly positive probability by player $i$ in $\sigma_i$.

\begin{thm}
In the secure localization game, no equilibrium strategy $\sigma=(\sigma_{\mathbf{v}},\sigma_{\mathbf{m}})$ can have $|supp(\sigma_i)|\in \mathbb{N}$ (i.e., $supp(\sigma_i)$ is a continuous space). \label{thm:spazioinfinito}
\end{thm}

\emph{Proof}. A necessary and sufficient condition such that a game with continuous actions admits an equilibrium where players randomize over a finite number of actions is that the continuous variables in the players' utility functions are separable, i.e., the utility functions can be expressed as the product of terms composed of only sum of variables. This does not hold in our case.\hfill$\Box$

We consider the situation in which the orthocenter $W$ of the triangle constituted of the three verifiers is a given data. By Theorem~\ref{thm:triangoloequilatero}, we know that the optimal verifiers' configuration is the equilateral triangle with edge's length equal to $R$. Consider the polar coordinate system with pole in the orthocenter $W$. Call $\alpha$ the angle between the polar axis and the line connecting a vertex $V_i$ to $W$. Since the verifiers must form an equilateral triangle and the verifiers have distance equal to $R$ from the pole, the verifiers' strategy can be compactly represented as a probability density over $\alpha$. Instead, the malicious node's strategy can be represented as a probability density over $U$ and $P$. We can show that the players' equilibrium strategies are the following.

\begin{thm} The players' equilibrium strategies are:
\begin{eqnarray*}
\sigma_{\mathbf{v}}^*  & = & \alpha \textnormal{ uniformly drawn from } [0,\frac{2\pi}{3}]\\
\sigma_{\mathbf{m}}^* & = & 
\begin{cases}
U & 
\begin{cases}
\rho_U = & 0.1394 R\\
\theta_U = &\textnormal{uniformly drawn from } [0,2 \pi]
\end{cases} \\
P & 
\begin{cases}
\rho_P = & 0.4286 R\\
\theta_P = & \theta_U + 0.2952
\end{cases}
\end{cases}
\end{eqnarray*}
\noindent and the expected utility of the malicious node is $0.001 R$.
\end{thm}

\emph{Proof}. By Theorem~\ref{thm:spazioinfinito}, the players must randomize over a continuous space of actions. We consider the verifiers' strategy. Easily, for symmetry reasons, the verifiers must randomize uniformly over all the possible values of $\alpha$. In particular, we can safely limit the randomization over $[0,2/3 \pi]$. We consider the malicious node's strategy. For symmetry, it randomize such that $\theta_u$ is uniformly drawn from $[0,2\pi]$. In order to compute the optimal $\rho_U$ and the polar coordinates of $P$, we solve the following optimization problem. We fix a value for $\theta_u$ and we search for the values of $\rho_U, \rho_P, \theta_P$ such that the malicious node's expected utility is the maximum one.
\begin{equation}
\max_{\rho_U,\rho_p,\theta_P} \int_{0}^{\frac{2\pi}{3}}\frac{u_{\mathbf{m}}}{\frac{2\pi}{3}}d\alpha
\end{equation}
The above optimization problem is non-linear. We solved it by discretizing the value of $\alpha$ with a step of $10^{-3}$ and by using conjugated subgradients. The result is the strategy reported above.\hfill$\Box$

Notice that, the expected utility of the malicious node drastically decreases with respect to the situation in which the strategy of the verifiers is pure, as it is $0.001 R$ with mixed strategy vs. $0.25 R$ with pure strategies. This is because with mixed strategies, the probability that the malicious node is not marked as robust or malicious is very small. Therefore, randomization over their strategies aids the verifiers to increase their expected utility and VM with mixed strategies can be considered to be robust.



\section{Conclusion}
\label{sec:conclusion}
The knowledge about the security of wireless sensor node localization
information is a fundamental challenge in order to provide trust
applications and data. Verifiable Multilateration is a secure
localization algorithm that defines two tests for evaluating node
behavior as malicious, or robust or in the worst case as unknown.  In
case of unknown node, VM does not have enough information for
evaluating the trustworthiness of the node. This lack of information
may be exploited by malicious user.  In this paper, in order to
improve the knowledge about the secure localization behavior VM has
been modelled as game, by means of game theory concepts. In fact a
verifier is the first player, while a malicious node is the second
player. Particularly we have analyzed the behavior in case of the
adoption of both a pure strategy and a mixed one.  The conducted
analysis demonstrates that, when the verifiers play a pure strategy,
the malicious node can always masquerade as unknown with a probability
of one and the deception is not negligible. When the verifiers play
mixed strategies, the malicious node can masquerade as unknown with a
very low probability and the expected deception is negligible. In the
future, we shall consider situations where a malicious attacker can
manipulate more nodes.

\section*{Acknowledgment}

This research has been partially funded by the European Commission,
Programme IDEAS-ERC, Project 227977-SMScom.



\bibliographystyle{IEEEtran}
\bibliography{wsngt}

\end{document}